\documentclass[12pt]{article}  
\usepackage{amsmath}
\usepackage{amssymb}                                          
\usepackage{graphicx}

\usepackage{mathtext}
\usepackage{epsf}
\usepackage{amsfonts}
\usepackage{gensymb}
\usepackage{float}

\def\apjl{ApJL}
\def\apjs{ApJ Suppl. Ser.}

\def\aap{A\&A}

\begin{document}


\title{On the possibility of registering X-ray flares related to fast radio bursts with the SRG/eROSITA telescope}

\author{A.D. Khokhryakova$^1$, D.A. Lyapina$^1$, S.B. Popov$^{1,2,3}$\\
$^1$ Department of Physics, Lomonosov Moscow State University\\
$^2$ Sternberg Astronomical Institute, Lomonosov Moscow State University\\
$^3$ Higher School of Economics, Moscow\\
}

\maketitle

\section*{Abstract}
    
In this note we discuss the possibility of detecting the accompanying  X-ray emission from  sources of fast radio bursts with the eROSITA telescope onboard the Spektr-RG observatory. 
It is shown that during  four years of the survey program about 300 bursts are expected to appear in the field of view of eROSITA. 
About 1\% of them will be detected by ground-based radio telescopes. 
For a total energy release $\sim~10^{46}$~ergs depending on the spectral parameters and absorption in the interstellar and intergalactic media, an X-ray flare can be detected from distances from $\sim 1$~Mpc (thermal spectrum with $kT=200$~keV and strong absorption) up to $\sim1$~Gpc (power-law spectrum with photon index $\Gamma=2 $ and realistic absorption).
Thus, eROSITA observations might help to provide important constraints on  parameters of sources of fast radio bursts, or may even allow to identify the X-ray transient counterparts, which will help to constrain models of fast radio bursts generation.

\section{Introduction}

Fast radio bursts (FRBs) are short ($\sim$ms) bright (peak fluxes up to $\sim$100~Jy) radio flashes (for a review, see \cite{Popov:2018}).\footnote{On-line arXiv:1806.03628.} The first event from this class of transients was introduced in 2007 in \cite{lorimer2007}.
Since then, several dozens of such bursts have been detected \cite{petroff2016}. \footnote{See the online FRB catalogue  http://www.frbcat.org.}  

A large dispersion measure and other considerations suggest an extragalactic origin of this phenomenon. By now, reliable identification has been made only for a single source of repeating fast radio bursts --- FRB 121102 (the notation: year-month-day)\footnote{Since this paper was accepted a new repeating source of FRBs has been discovered (The CHIME/FRB Collaboration 2019).}. 
The source is located in a dwarf galaxy with active star formation at redshift $z=0.193$ (corresponding photometric distance is 972 Mpc)  \cite{tendulkar2017}.

Nowadays, there are lots of models explaining the nature of FRB sources (see catalogue of theories in \cite{platts2018}).
This reflects a wide uncertainty in the description of the nature of these objects.
Nonetheless, the main current models associate FRB with neutron stars.

Magnetic energy dissipation in young neutron stars is one of the most prospective hypotheses about the nature of FRBs. Particularly, FRB generation may be associated with magnetar hyperflares (see review \cite{2015RPPh...78k6901T} for this type of sources and forms of their activity). This model was proposed immediately after the discovery of the first FRB \cite{popov2007}\footnote{Originally this paper was published just as an e-print. It can be found at arXiv.org: 0710.2006.}.
An important prediction of this model is a simultaneous radiation pulse from FRBs in X-rays and, possibly, gamma-ray ranges (see, for example, \cite{lyubarsky2014} and \cite{murase2016}). The luminosity of the only observed hyperflare from SGR 1806-20  was $\sim10^{47}~$erg s$^{-1}$, and the total energy release was $\sim10^{46}$~erg \cite{2005Natur.434.1107P}.

In this paper we discuss the possibility of registration of X-ray radiation accompanying FRBs with the eROSITA telescope, the detailed description of which is presented in \cite{merloni2012}.
Obtaining a positive result in such observations will give an opportunity to verify (or derive strict limits) the hyperflare  model of FRBs.

eROSITA (extended ROentgen Survey with an Imaging Telescope Array) is the primary instrument on the forthcoming Spectrum-Roentgen-Gamma (SRG) mission. 
This telescope will be used to study the entire sky in the X-ray band. Over 4 years of operation eROSITA will make 8 full surveys of the sky in the energy range from a few tenths up to 10 keV.

In the soft X-ray band ($\sim$0.5-2 keV) this instrument will be about 20
times more sensitive than the ROSAT satellite. In the hard band (2-10 keV) it will provide the first imaging survey of the whole sky at those energies.

\section{Estimating the number of bursts}

In this section we estimate the number of FRB flares that will fall in the field of view of eROSITA.
To date, various estimates of the number of FRBs qualitatively converge with each other, giving a rate of about several thousand events per day in the entire sky with a flux of more than a few tenths of Jy. In the estimates below, we will use the value of $N_{\Sigma}=10^4$ bursts per day, which matches the analysis carried out in \cite{thornton2013}, \cite{vander2016}.

Given that the field of view of eROSITA is 0.833 square degrees,
it can be calculated that approximately 0.2 bursts are seen per day, which can potentially be detected by ground-based radio telescopes. On average, one burst will fall into the field of view of the telescope in 5 days, and in 4 years of survey observations the number of such events will be about 300.

It should be noted that a comparable number of bursts should have fallen within the XMM-Newton field of view in $\sim18$ years of operation. However, it is difficult to identify a short non-repeating weak burst. Due to the small number of FRBs registered in the radio band, there were no cases where XMM-Newton or another instrument would have observed the radio burst area at the time of the event. This is primarily due to the low rate of registration of radio bursts during the work of XMM-Newton. In the next few years, the number of detected bursts will increase significantly. It is therefore essential to estimate the number of future detected radio bursts that fall within the field of view of eROSITA.

Currently, both radio telescopes already operating for a long time (64-m Parkes telescope, the Arecibo antenna, Green Bank Telescope) and new instruments such as ASKAP (Australian Square Kilometre Array Pathfinder) and UTMOST (Molonglo Observatory Synthesis Telescope) are actively used to search for FRBs.  In addition, the search will be conducted with the new 500-m FAST antenna in China. It is expected that in the near future HIRAX (Hydrogen Intensity Real-time Analysis eXperiment) and CHIME (Canadian Hydrogen Intensity Mapping Experiment) will be able to detect several dozen of flashes per day \cite{rajwade2017}. Optimistic estimates raise this number to one hundred bursts per day for each telescope \cite{walters2018}.

Therefore, a reasonable assumption is that in total ground-based radio telescopes will detect $\sim$1\% of all bursts. 
If bursts occur evenly across the sky (which is a good assumption due to the extragalactic origin of FRBs and the absence of correlation with known local extragalactic structures), it can be calculated that the number of detected radio bursts falling into the field of view of eROSITA will be
\begin{equation}
N = \frac{N_X N_R}{N_{\Sigma}},
\end{equation}
where $N_R$ is the number of flashes detected by ground-based telescopes, $N_X$ is the number of flashes detected by eROSITA. Hence, it can be estimated that, for the total of all the time of survey observations on the SRG satellite, several events ($\sim3$) recorded by ground-based radio telescopes (HIRAX, CHIME, ASKAP, UTMOST, etc.) will fall in the field of view of the eROSITA.\footnote{
In addition, one can expect to have approximately one of the FRB recorded by radio telescopes within the ART-XC field of view during the time of the satellite operation.} 
This makes it relevant to conduct a more detailed evaluations of the ability of eROSITA to register X-ray flares that may accompany the radio bursts.

\section{The possibility of recording hyperflares}

In this section we consider parameters of X-ray flares in order to evaluate whether the sensitivity of eROSITA is high enough for their registration.

The duration of the main peak of giant flares and hyperflares from magnetars is $\gtrsim0.1-0.2$ s \cite{1999Natur.397...41H, 2005Natur.434.1098H}.  This exceeds the nominal integration time for eROSITA, which is 50 ms. Thus, although some of the incoming photons may not be registered separately, it still can be expected that the hyperflare will lead to several counts on the detector.

We accept that for reliable signal detection, the telescope must register 10 photons from the source for the entire time of the X-ray flash.\footnote{
If we talk about the presence of a radio trigger, i.e. if the arrival time and the coordinates of the flare are known, then the criterion of 10 photons can be significantly softened to 2-3 photons from a compact area corresponding to the angular resolution of the telescope. 
On the other hand, the finite integration time (50 ms) of the detector can lead to the situation when those $\sim$10 photons, which arrived during the 100-200 ms flash, will ``overlap'' each other (the pile-up effect) in individual pixels, reducing the number of actually recorded events. Thus, our criterion of 10 arrived photons seems reasonable.}
Assuming that the telescope detected 10 photons, we construct the dependence of the energy release of the source from the distance for several spectral models.

We will examine several options representing a wide range of X-ray burst spectra that can potentially accompany FRBs (primarily those that can correspond to magnetar hyperflares). These are blackbody spectra for temperatures $kT = 30$ keV and $kT = 200$ keV and power-law spectra with photon indexes $\Gamma = 0.5$ and $\Gamma = 2$.

As long as the observations are carried out in a fairly soft part of the X-ray spectrum, the photon flux will be significantly attenuated due to interstellar absorption 
(in the interstellar medium of the Galaxy, in the intergalactic medium and in the interstellar medium of the source's host galaxy).  
At a column density of hydrogen atoms $N_H$ the flux weakens by a factor of $e^{-\sigma N_H} $. To calculate $\sigma$, we use data from \cite{morrison1983}:
\begin{equation}
\sigma=\frac{1}{E}C_2+\frac{1}{E^2}C_1+\frac{1}{E^3}C_0,
\end{equation}
where $E$ is the photon energy and the coefficients $C_0, C_1, C_2$ were taken from the abovementioned paper. Given the evaluative nature of our work, and that we are interested only in the total energy release in a quite wide X-ray band, the non-usage of more detailed results on interstellar absorption (see, for example, \cite{wilms2000}) is not crucial for our purposes. Note also that in the case of FRBs, a significant part of the absorption should be unrelated to the matter inside our Galaxy, so accurate calculations of the $\sigma$ parameter become even more uncertain.

Also, we take into consideration the dependence of the effective area of the telescope on the wavelength (see Fig. 1). 
Several of  eROSITA mirror systems will be covered by filters that cut off the soft part of the spectrum. We use data on the effective area assuming that five systems out of seven are covered by such filters.
Relevant data are taken from the site https://wiki.mpe.mpg.de/eRosita.

\begin{figure}
\includegraphics[scale=0.8]{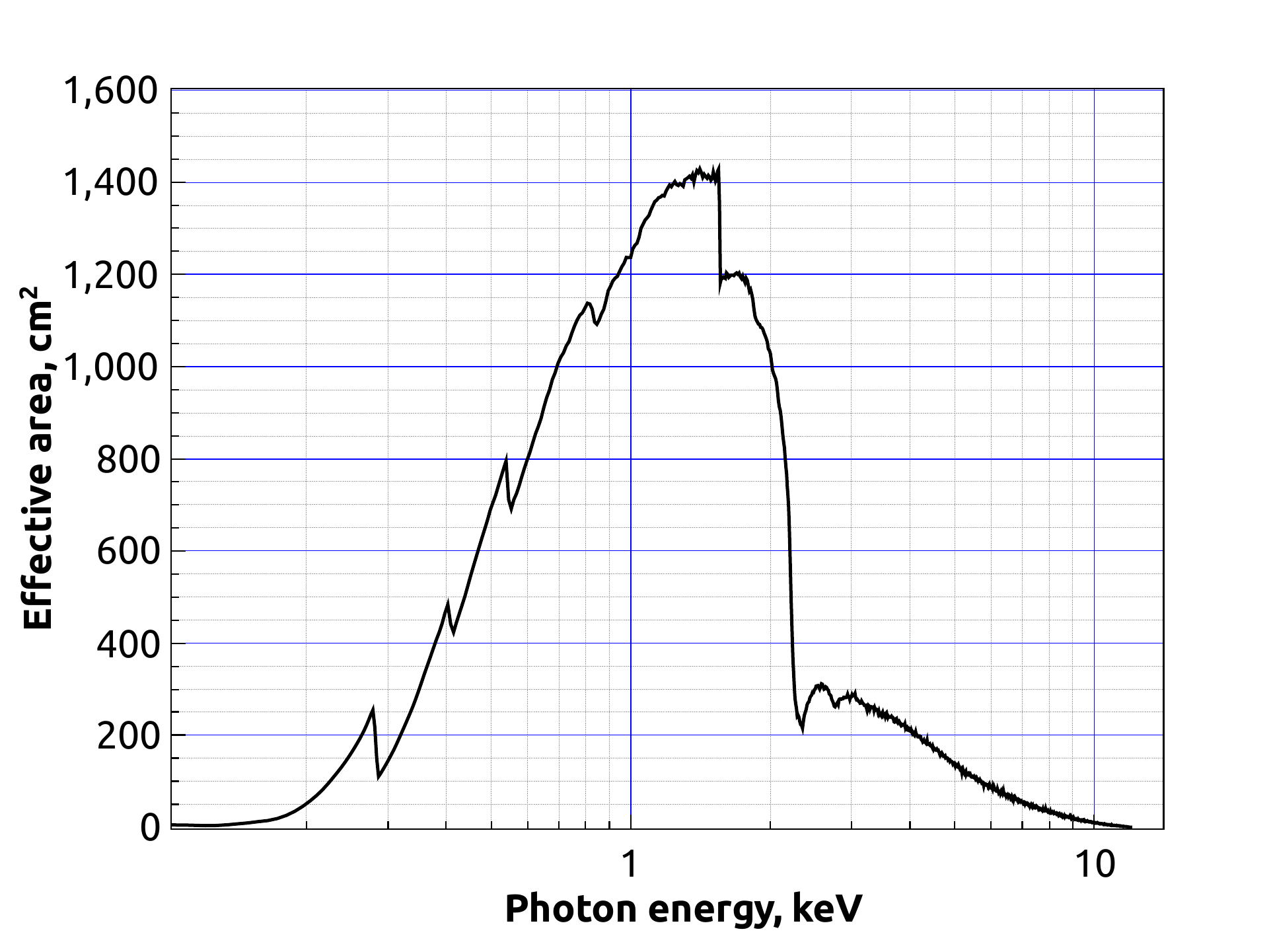}
\caption{Effective area of the eROSITA telescope versus photon energy. The data are presented under the assumption that five out of  seven mirror systems are covered with filters cutting off the softest part of the X-ray spectrum.}
\end{figure}

\subsection{Power-law spectrum}

Models of X-ray flares with power-law spectrum in the application to FRBs have recently been considered in \cite{scholz2017}. 
We suggest that the spectrum of an object is given by the equation
\begin{equation}
dN = C E^{ - \Gamma}  e^{-(E/E_{cutoff})} dE,
\end{equation}
where $E_{cutoff} = 500$ keV is the cutoff energy of the spectrum, $C$ is the dimensional constant determined from the normalization of the total energy release to $10^{47}$ erg.

Here and below, $E$~is the photon energy in keV.
The value of $\Gamma$ is taken to be 0.5 and 2 as extreme cases. 
For $\Gamma = 0.5 $ we get $C = 1.01 \times 10^{43}$~erg$^{-1/2}$. 
At $\Gamma = 2$, the integral diverges at $E \rightarrow 0$, so it is necessary to choose a nonzero lower limit of integration. We vary it from $10^{-5}$ to 0.1 keV, and it turns out that $C$ changes by no more than an order of magnitude, which does not significantly affect further estimates. The lower limit is chosen to be $0.001$ keV, with $C = 9.7 \times 10^{45}$~erg. This value is used in subsequent estimations.

Taking into account absorption and  dependence of the eROSITA effective area on the photon energy, the radiation energy arriving at the detector from the source at a distance of $r$ (fluence multiplied by the effective area of the detector) will be:
\begin{equation}
F_d = \int_{E_1}^{E_2} {\frac{C E_{}^{1 - \Gamma} e^{-E_{}/E_{cutoff}}  e^{- \sigma N_H}  S_{eff}(E_{})  dE_{}} {4 \pi  r^2}}.
\end{equation}

Here we neglect the redshift effect because the distances for potentially detectable flares do not exceed 1 Gpc, which corresponds to $z<0.2$. In addition, it is important to emphasize the approximate nature of our estimates (for example, due to the uncertainties with absorption on the line of sight).

The number of detected photons is
\begin{equation}
N_d = \int_{E_1}^{E_2} {\frac{C E_{}^{- \Gamma} e^{-E_{}/E_{cutoff}}  e^{- \sigma N_H}  S_{eff}(E_{})  dE} {4 \pi  r^2}}.
\label{r10}
\end{equation}
An example for a distance of 100 Mpc and a flare energy of $10^{47}$~erg is given in Table 1 for several spectral models and column densities.

\begin{table*}[t]
\caption{Number of registered photons for a burst at the distance 100 Mpc with the energy   $10^{47}$~erg.}
\label{tabular:timesandtenses}
\begin{center}
\begin{tabular}{l l l l l}
\hline
$N_H$, cm$^{-2}$ & $kT = 30$ keV  & $kT = 200$ keV & $\Gamma = 0.5$ & $\Gamma = 2$\\
\hline \hline
0 & 1.9 & 0.017 & 12.6 & $13000$ \\
$10^{22}$ & 1.35 & 0.012 & 4.8 & $1600$\\
$10^{24}$ & 0.087 & 0.0008 & 0.087 & 4.7 \\
\hline
\end{tabular}
\end{center}
\end{table*}

Assuming that 10 photons were registered from the flare, we can calculate the dependence of the total energy of the flare on the distance.
The distance $r_{10}$, from which 10 photons come from a flash with an energy of $10^{47}$~erg, can be determined from the equation (\ref{r10}). 
Next, the total energy is $E_{total}=10^{47} (r/r_{10})^2$ erg. 
Here we consider $N_H$ to be a fixed parameter, i.e. a change in $r$ does not lead to a change in absorption, so the quadratic dependence is preserved. 
This simplification is possible due to the large uncertainty of the parameter $N_H$.

\begin{figure}
\includegraphics[scale=0.7]{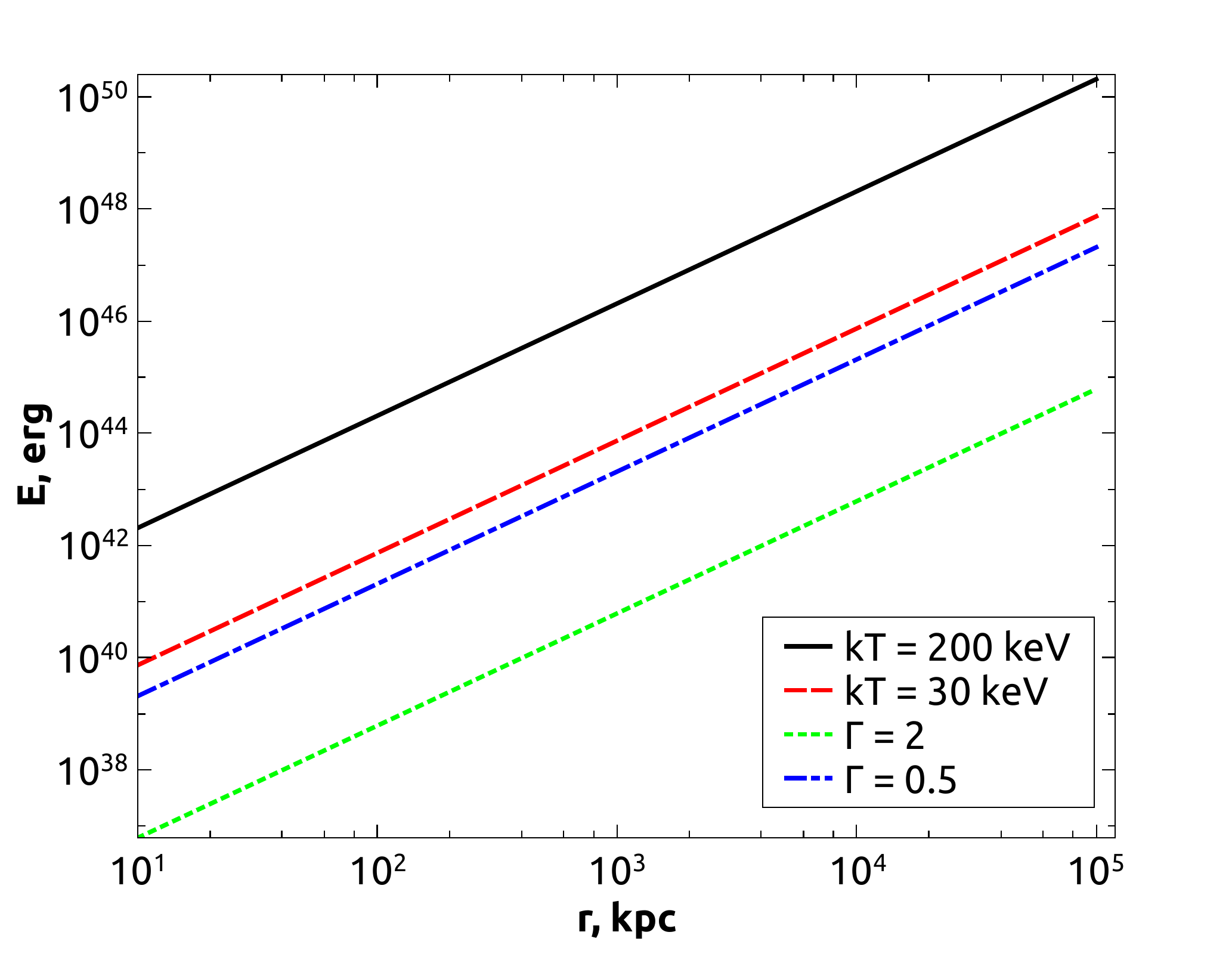}
\caption{Dependence of the burst energy on distance for different models of the hyperflares spectra under the assumption that 10 photons are registered. The number of hydrogen atoms on the line of sight is taken to be $N_H = 10^{22}$cm$^{-2}$.}
\end{figure}

\begin{figure}
\includegraphics[scale=0.6]{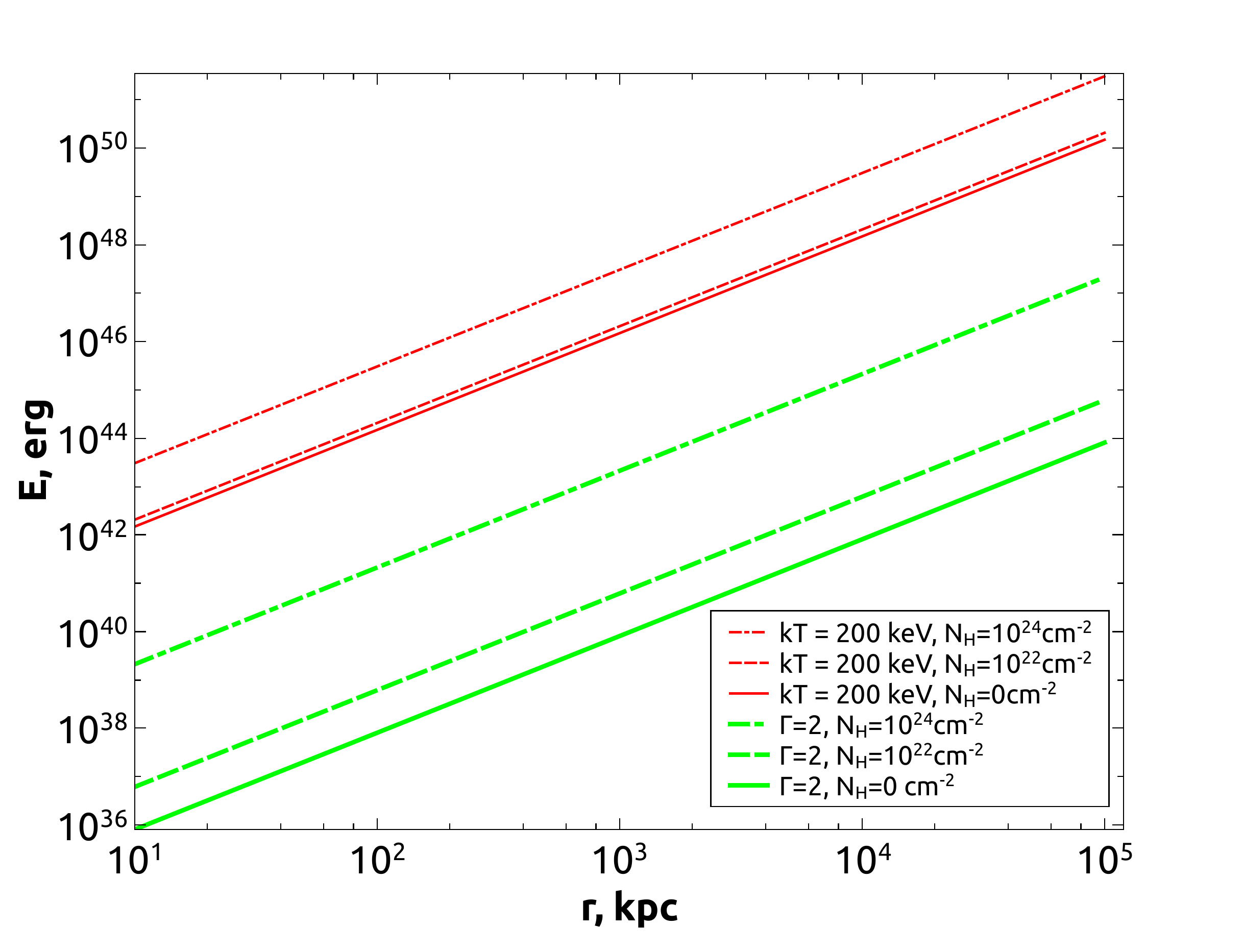}
\caption{Dependence of the burst energy on  distance for power-law (thick green lines) and thermal spectra (thin red lines) under the assumption that 10 photons are registered. Solid lines correspond to $N_H=0$, dashed --- to $N_H=10^{22}$cm$^{-2}$,  and dot-dashed --- to $N_H=10^{24}$cm$^{-2}$.
}
\end{figure}

\subsection{Blackbody spectrum}

 Similar calculations are made also with the assumption that the spectrum of the X-ray burst is thermal. Observations show that the  spectra of giant magnetar flares can be well described by a blackbody spectrum with a temperature from $\sim$30 to $\sim200$ keV (see, for example, \cite{2018arXiv180305716E} and references therein). 
 The X-ray radiation is associated with the fireball retained by the magnetosphere with a characteristic size of a few hundred kilometers.\footnote{
 Certainly, the radio emission corresponding to FRBs is non-thermal (and coherent) and is generated, apparently, in another spatial domain, for instance, in a shell similar to the pulsar nebula surrounding magnetars (see \cite{lyubarsky2014}). Several Galactic magnetars and high-B pulsars have such shells.}

In the case of such a spectrum, we have
\begin{equation}
    B_E(T, E) = \frac{2  }{c^2 h^3} \frac {E_{}^3}  {e^{\frac{E_{}} {k T}} - 1},
\end{equation}
\begin{equation}
F_d = \int_{E_1}^{E_2} {\frac{ C_p \pi B_E (T,E)   e^{- \sigma N_H}  S_{eff}(E_{}) dE} {4 \pi r^2}},
\end{equation}
where $C_p$ can be found from the normalization to the total energy release $10^{47}$ erg.
Number of detected photons is:
\begin{equation}
N_d = \int _{E_1}^{E_2}{\frac{ C_p \pi B_E (T,E) E^{-1} e^{- \sigma N_H}  S_{eff}(E_{}) dE} {4 \pi r^2}}.
\end{equation}


The results for all the considered spectra are shown in Figs.~2 and 3 for different values of the column density $N_H$.

Note that flares can also be registered by the ART-XC telescope.
Preliminary estimates show that for a thermal spectrum with $kT=30$~keV ($N_H=10^{22}$~cm$^{-2}$) one can expect several counts in the case of 
a flare with an energy of $\sim 10^{46}$~erg from distances of the order of tens of Mpc.
Nevertheless, as long as the field of view of the ART-XC is about three times smaller than that of eROSITA, the expected number of potentially detectable events decreases, correspondingly.

\section{Discussion}

Evaluation of distances to the FRBs show that if the intergalactic medium (rather than the matter in the immediate neighborhood of the source) dominates in the dispersion measure, then typical distances might be about 1~Gpc, and the minimum distances might be about 100 Mpc (see, e.g., the dispersion measure estimates in the intergalactic medium for FRBs detected by the ASKAP \cite{shannon2018}).
This means that most X-ray flares are potentially can be recorded with eROSITA only for soft power-law spectra and a sufficiently large energy release.
However, it is essential that there are alternative models in which a significant part of the FRB dispersion measure is gathered in the environment near the source (see, for example, \cite{2016MNRAS.462..941L} and the references given there).  
In this instance, the average distance will be smaller.


Note, however, that in \cite{2016MNRAS.462..941L} the authors consider the pulsar model, not the magnetar one. In that model,  no significant X-ray emission is expected. On the other hand, in the scenario of magnetar hyperflares, the interstellar medium around the source  most probably will be insufficiently dense to make a significant contribution to the dispersion measure. Thus, it is unlikely that if FRBs are relatively close ($\lesssim100$~Mpc), the radio burst will be accompanied by a powerful X-ray flare.

In this paper, we did not examine the option with a soft thermal spectrum $kT=10$~keV used, for example, in \cite{scholz2017}. 
This is due to the fact that such a soft spectrum must be atypical for hyperflares from magnetars. 
For example, in the case of the SGR 1806-20 hyperflare, the spectrum can be described by a blackbody with a temperature of $\sim$200 keV \cite{2005Natur.434.1098H}. Surely, for a softer spectrum and the same total energy release, the flux in the eROSITA band increases, and the flare will be detectable from distances $\gtrsim 100$ Mpc.

Note that  calculations presented above are also applicable for estimates of the detectability of hyperflares from extragalactic magnetars unrelated to any possible connection with fast radio bursts. Certainly, in the vast majority of cases it will be difficult to identify a faint flare as an event associated with the activity of a distant magnetar. However, with good localization, allowing to identify the host galaxy, and the presence of repetitions, it will be possible uncover the nature of such event.

It should also be noted that when the review program on the Spectrum-Roentgen-Gamma satellite is finished, it will be possible to perform long-term simultaneous observations with the eROSITA and ART-XC instruments and radio telescopes. 
This will be especially important in case of detection of sources of repeating bursts at distances less than a few hundred Mpc.

\section{Conclusions}

We examined the possibility of detecting the hyperflares accompanying the FRB emission with the eROSITA telescope. We have shown that approximately one hyperflare per year can appear within the eROSITA field of view simultaneously with the registration of the FRB by ground-based radio telescopes. At the same time, the sensitivity of the X-ray telescope turned out to be quite sufficient to detect a hyperflare with an energy of $10^{46}$ erg from distances about tens or even hundreds of Mpc for realistic spectral parameters.

\section*{Acknowledgements}

We are grateful to the referee for several useful comments that helped to improve the paper.
We thank prof. N.I. Shakura and dr. K.L. Malanchev for their comments.
A.D. Khokhryakova and D.A. Lyapina thank the Moscow State University Development Program in the nomination ``Outstanding Scientific Schools of the Moscow State University''.
The work of S.B. Popov was supported by the RSF grant no. 19-12-00084.

\vskip 0.1cm

This is an authors' version of translation of the paper which appeared in Astronomy Letters (in Russian: Pis'ma v Astronomicheskij Zhurnal) in 2019 (N3). 

\bibliography{main}

\begin{thebibliography}{10}

\bibitem{Popov:2018}
S.~B. Popov, K.~A. Postnov, and M.~S. Pshirkov.
\newblock Fast radio bursts.
\newblock {\em Physics Uspekhi}, 61(10):965--979, 2018.

\bibitem{lorimer2007}
D.~R. {Lorimer}, M.~{Bailes}, M.~A. {McLaughlin}, D.~J. {Narkevic}, and
  F.~{Crawford}.
\newblock {A Bright Millisecond Radio Burst of Extragalactic Origin}.
\newblock {\em Science}, 318:777, November 2007.

\bibitem{petroff2016}
E.~{Petroff}, E.~D. {Barr}, A.~{Jameson}, E.~F. {Keane}, M.~{Bailes},
  M.~{Kramer}, V.~{Morello}, D.~{Tabbara}, and W.~{van Straten}.
\newblock {FRBCAT: The Fast Radio Burst Catalogue}.
\newblock {\em \pasa}, 33:e045, September 2016.

\bibitem{tendulkar2017}
S.~P. {Tendulkar}, C.~G. {Bassa}, J.~M. {Cordes}, G.~C. {Bower}, C.~J. {Law},
  S.~{Chatterjee}, E.~A.~K. {Adams}, S.~{Bogdanov}, S.~{Burke-Spolaor}, B.~J.
  {Butler}, P.~{Demorest}, J.~W.~T. {Hessels}, V.~M. {Kaspi}, T.~J.~W. {Lazio},
  N.~{Maddox}, B.~{Marcote}, M.~A. {McLaughlin}, Z.~{Paragi}, S.~M. {Ransom},
  P.~{Scholz}, A.~{Seymour}, L.~G. {Spitler}, H.~J. {van Langevelde}, and R.~S.
  {Wharton}.
\newblock {The Host Galaxy and Redshift of the Repeating Fast Radio Burst FRB
  121102}.
\newblock {\em \apjl}, 834:L7, January 2017.

\bibitem{platts2018}
E.~{Platts}, A.~{Weltman}, A.~{Walters}, S.~P. {Tendulkar}, J.~E.~B. {Gordin},
  and S.~{Kandhai}.
\newblock {A Living Theory Catalogue for Fast Radio Bursts}.
\newblock {\em ArXiv e-prints: 1810.05836}, October 2018.

\bibitem{2015RPPh...78k6901T}
R.~{Turolla}, S.~{Zane}, and A.~L. {Watts}.
\newblock {Magnetars: the physics behind observations. A review}.
\newblock {\em Reports on Progress in Physics}, 78(11):116901, November 2015.

\bibitem{popov2007}
S.~B. {Popov} and K.~A. {Postnov}.
\newblock {Hyperflares of SGRs as an engine for millisecond extragalactic radio
  bursts}.
\newblock In H.~A. {Harutyunian}, A.~M. {Mickaelian}, and Y.~{Terzian},
  editors, {\em Evolution of Cosmic Objects through their Physical Activity},
  pages 129--132, November 2010.

\bibitem{lyubarsky2014}
Y.~{Lyubarsky}.
\newblock {A model for fast extragalactic radio bursts}.
\newblock {\em \mnras}, 442:L9--L13, July 2014.

\bibitem{murase2016}
K.~{Murase}, K.~{Kashiyama}, and P.~{M{\'e}sz{\'a}ros}.
\newblock {A burst in a wind bubble and the impact on baryonic ejecta:
  high-energy gamma-ray flashes and afterglows from fast radio bursts and
  pulsar-driven supernova remnants}.
\newblock {\em \mnras}, 461:1498--1511, September 2016.

\bibitem{2005Natur.434.1107P}
D.~M. {Palmer}, S.~{Barthelmy}, N.~{Gehrels}, R.~M. {Kippen}, T.~{Cayton},
  C.~{Kouveliotou}, D.~{Eichler}, R.~A.~M.~J. {Wijers}, P.~M. {Woods},
  J.~{Granot}, Y.~E. {Lyubarsky}, E.~{Ramirez-Ruiz}, L.~{Barbier},
  M.~{Chester}, J.~{Cummings}, E.~E. {Fenimore}, M.~H. {Finger}, B.~M.
  {Gaensler}, D.~{Hullinger}, H.~{Krimm}, C.~B. {Markwardt}, J.~A. {Nousek},
  A.~{Parsons}, S.~{Patel}, T.~{Sakamoto}, G.~{Sato}, M.~{Suzuki}, and
  J.~{Tueller}.
\newblock {A giant {$\gamma$}-ray flare from the magnetar SGR 1806 - 20}.
\newblock {\em \nat}, 434:1107--1109, April 2005.

\bibitem{merloni2012}
A.~{Merloni}, P.~{Predehl}, W.~{Becker}, H.~{B{\"o}hringer}, T.~{Boller},
  H.~{Brunner}, M.~{Brusa}, K.~{Dennerl}, M.~{Freyberg}, P.~{Friedrich},
  A.~{Georgakakis}, F.~{Haberl}, G.~{Hasinger}, N.~{Meidinger}, J.~{Mohr},
  K.~{Nandra}, A.~{Rau}, T.~H. {Reiprich}, J.~{Robrade}, M.~{Salvato},
  A.~{Santangelo}, M.~{Sasaki}, A.~{Schwope}, J.~{Wilms}, and t.~{German
  eROSITA Consortium}.
\newblock {eROSITA Science Book: Mapping the Structure of the Energetic
  Universe}.
\newblock {\em ArXiv e-prints: 1209.3114}, September 2012.

\bibitem{thornton2013}
D.~{Thornton}, B.~{Stappers}, M.~{Bailes}, B.~{Barsdell}, S.~{Bates}, N.~D.~R.
  {Bhat}, M.~{Burgay}, S.~{Burke-Spolaor}, D.~J. {Champion}, P.~{Coster},
  N.~{D'Amico}, A.~{Jameson}, S.~{Johnston}, M.~{Keith}, M.~{Kramer},
  L.~{Levin}, S.~{Milia}, C.~{Ng}, A.~{Possenti}, and W.~{van Straten}.
\newblock {A Population of Fast Radio Bursts at Cosmological Distances}.
\newblock {\em Science}, 341:53--56, July 2013.

\bibitem{vander2016}
S.~{Vander Wiel}, S.~{Burke-Spolaor}, E.~{Lawrence}, C.~J. {Law}, and G.~C.
  {Bower}.
\newblock {Rare Event Statistics Applied to Fast Radio Bursts}.
\newblock {\em ArXiv e-prints}, December 2016.

\bibitem{rajwade2017}
K.~M. {Rajwade} and D.~R. {Lorimer}.
\newblock {Detecting fast radio bursts at decametric wavelengths}.
\newblock {\em \mnras}, 465:2286--2293, February 2017.

\bibitem{walters2018}
A.~{Walters}, A.~{Weltman}, B.~M. {Gaensler}, Y.-Z. {Ma}, and A.~{Witzemann}.
\newblock {Future Cosmological Constraints From Fast Radio Bursts}.
\newblock {\em \apj}, 856:65, March 2018.

\bibitem{1999Natur.397...41H}
K.~{Hurley}, T.~{Cline}, E.~{Mazets}, S.~{Barthelmy}, P.~{Butterworth},
  F.~{Marshall}, D.~{Palmer}, R.~{Aptekar}, S.~{Golenetskii}, V.~{Il'Inskii},
  D.~{Frederiks}, J.~{McTiernan}, R.~{Gold}, and J.~{Trombka}.
\newblock {A giant periodic flare from the soft {$\gamma$}-ray repeater
  SGR1900+14}.
\newblock {\em \nat}, 397:41--43, January 1999.

\bibitem{2005Natur.434.1098H}
K.~{Hurley}, S.~E. {Boggs}, D.~M. {Smith}, R.~C. {Duncan}, R.~{Lin},
  A.~{Zoglauer}, S.~{Krucker}, G.~{Hurford}, H.~{Hudson}, C.~{Wigger},
  W.~{Hajdas}, C.~{Thompson}, I.~{Mitrofanov}, A.~{Sanin}, W.~{Boynton},
  C.~{Fellows}, A.~{von Kienlin}, G.~{Lichti}, A.~{Rau}, and T.~{Cline}.
\newblock {An exceptionally bright flare from SGR 1806-20 and the origins of
  short-duration {$\gamma$}-ray bursts}.
\newblock {\em \nat}, 434:1098--1103, April 2005.

\bibitem{morrison1983}
R.~{Morrison} and D.~{McCammon}.
\newblock {Interstellar photoelectric absorption cross sections, 0.03-10 keV}.
\newblock {\em \apj}, 270:119--122, July 1983.

\bibitem{wilms2000}
J.~{Wilms}, A.~{Allen}, and R.~{McCray}.
\newblock {On the Absorption of X-Rays in the Interstellar Medium}.
\newblock {\em \apj}, 542:914--924, October 2000.

\bibitem{scholz2017}
P.~{Scholz}, S.~{Bogdanov}, J.~W.~T. {Hessels}, R.~S. {Lynch}, L.~G. {Spitler},
  C.~G. {Bassa}, G.~C. {Bower}, S.~{Burke-Spolaor}, B.~J. {Butler},
  S.~{Chatterjee}, J.~M. {Cordes}, K.~{Gourdji}, V.~M. {Kaspi}, C.~J. {Law},
  B.~{Marcote}, M.~A. {McLaughlin}, D.~{Michilli}, Z.~{Paragi}, S.~M. {Ransom},
  A.~{Seymour}, S.~P. {Tendulkar}, and R.~S. {Wharton}.
\newblock {Simultaneous X-Ray, Gamma-Ray, and Radio Observations of the
  Repeating Fast Radio Burst FRB 121102}.
\newblock {\em \apj}, 846:80, September 2017.

\bibitem{2018arXiv180305716E}
P.~{Esposito}, N.~{Rea}, and G.~L. {Israel}.
\newblock {Magnetars: a short review and some sparse considerations}.
\newblock {\em ArXiv e-prints: 1803.05716}, March 2018.

\bibitem{shannon2018}
R.~M. {Shannon}, J.-P. {Macquart}, K.~W. {Bannister}, R.~D. {Ekers}, C.~W.
  {James}, S.~{Oslowski}, H.~{Qiu}, M.~{Sammons}, A.~W. {Hotan}, M.~A.
  {Voronkov}, R.~J. {Beresford}, M.~{Brothers}, A.~J. {Brown}, J.~D. {Bunton},
  A.~P. {Chippendale}, C.~{Haskins}, M.~{Leach}, M.~{Marquarding},
  D.~{McConnell}, M.~A. {Pilawa}, E.~M. {Sadler}, E.~R. {Troup}, J.~{Tuthill},
  M.~T. {Whiting}, J.~R. {Allison}, C.~S. {Anderson}, M.~E. {Bell}, J.~D.
  {Collier}, G.~{Gurkan}, G.~{Heald}, and C.~J. {Riseley}.
\newblock {The dispersion-brightness relation for fast radio bursts from a
  wide-field survey.}
\newblock {\em \nat}, 562:386--390, October 2018.

\bibitem{2016MNRAS.462..941L}
M.~{Lyutikov}, L.~{Burzawa}, and S.~B. {Popov}.
\newblock {Fast radio bursts as giant pulses from young rapidly rotating
  pulsars}.
\newblock {\em \mnras}, 462:941--950, October 2016.

\end{thebibliography}

\end{document}